\documentclass[
reprint,
amsmath,
amssymb,
aps,
prb,
floatfix,
]{revtex4-2}
\usepackage{graphicx}
\usepackage{hyperref}
\usepackage{xcolor}
\usepackage{multirow}

\begin{document}

\title{
Compositional phase stability in medium-entropy and high-entropy Cantor-Wu alloys from an \emph{ab initio} all-electron, Landau-type theory and atomistic modelling}

\author{Christopher D. Woodgate}
\email{christopher.woodgate@warwick.ac.uk}
\author{Julie B. Staunton}
\affiliation{Department of Physics, University of Warwick, Coventry, CV4 7AL, United Kingdom}

\begin{abstract}
We describe implementation and analysis of a first-principles theory, derived in an earlier work, for the leading terms in an expansion of a Gibbs free energy of a multi-component alloy in terms of order parameters that characterize potential, compositional phases. The theory includes effects of rearranging charge and other electronics from changing atomic occupancies on lattice sites. As well as the rigorous description of atomic short-range order in the homogeneously disordered phase, pairwise interaction parameters suited for atomistic modelling in a multicomponent setting can be calculated. From our study of an indicative series of the Cantor-Wu alloys, NiCo, NiCoCr, NiCoFeCr, and NiCoFeMnCr, we find that the interactions are not approximated well either as pseudobinary or restricted to nearest neighbour range. Our computed order-disorder transition temperatures are low, consistent with experimental observations, and the nature of the ordering is dominated by correlations between Ni, Co, and Cr, while Fe and Mn interact weakly. Further atomistic modelling suggests that there is no true single-phase low-temperature ground state for these multicomponent systems. Instead the single-phase solid solution is kept stable to low temperatures by the large configurational entropy and the Fe, Mn dilution effects. The computationally cost-effectiveness of our method makes it a good candidate for further exploration of the space of multicomponent alloys.

\end{abstract}

\maketitle

\section{Introduction}

With the emergence of the interest over the last ten years or so in high entropy alloys~\cite{cantor_microstructural_2004, yeh_nanostructured_2004,gao_high-entropy_2016, george_high-entropy_2019} the motivation to explore the vast composition space inhabited by multicomponent alloys has grown dramatically. 
While much work had been focused on alloys composed of one or two principal elements with small additions of other components to enhance desired properties, seminal papers by Cantor {\it et al}.~\cite{cantor_microstructural_2004} and Yeh {\it et al}.~\cite{yeh_nanostructured_2004} demonstrated that multiple elements with similar atomic sizes can be combined in roughly equal ratios to form a single phase solid solution. The stabilisation of the solid solution was attributed to the large contribution to the free energy from the configurational entropy, sometimes also referred to as the entropy of mixing. As a result, these alloys are referred to as `high-entropy' alloys (HEAs), although the terms `multicomponent' and `multi-principal element' are also often used. There is also a convention of referring to alloys with only three or four component species as `medium entropy'.
\par
Many of the multicomponent alloys discovered thus far possess exceptional physical properties for applications, but there still remains a huge, largely unexplored space of candidate materials. The guiding principles for designing these materials are also not well understood; a large number of candidate HEAs have been shown to fail to form single-phase solid solutions~\cite{wu_recovery_2014}. A key challenge for computational materials modelling, which starts out from an {\it ab initio} Density Functional Theory (DFT) basis, therefore, is to develop effective techniques for these systems to guide experiment, both by predicting phase diagrams and also the physical properties of candidate alloys for applications. These modelling techniques need to be computationally efficient and, in particular, should scale well with the number of component chemical species. Effective medium theories for coping with the necessity for averaging over many compositional configurations, such as the Coherent Potential Approximation (CPA)~\cite{soven_coherent-potential_1967}, are good candidates for satisfying both of these requirements~\cite{tian_alloying_2017,niu_first_2016,robarts_extreme_2020}. Studies of alloy phase behaviour using these techniques can analyse rigorously the stability of the high-temperature, high-symmetry disordered solid solution to chemical fluctuations using the concentration wave formalism~\cite{khachaturyan_ordering_1978, gyorffy_concentration_1983, staunton_compositional_1994, singh_atomic_2015, khan_statistical_2016, singh_tuning_2019, singh_first-principles_2020,singh_martensitic_2021, schonfeld_local_2019}, and it is this approach which we will take in this work.
\par
A variety of alternative approaches have been also used to study the onset and nature of compositional order in the high- and medium-entropy alloys, including {\it ab initio} DFT calculations on large supercells, molecular dynamics simulations, CALPHAD and semi-empirical calculations~\cite{ferrari_frontiers_2020,feng_design_2016,feng_phase_2018,widom_modeling_2018,sorkin_generalized_2020,ikeda_ab_2019, troparevsky_criteria_2015, feng_first-principles_2017, gao_thermodynamics_2017, gorsse_about_2018}. Some studies also use cluster expansions and machine-learning approaches to develop effective hamiltonians with which to perform atomistic modelling~\cite{pei_machine-learning_2020, huang_machine-learning_2019,fernandez-caballero_short-range_2017,liu_monte_2021}.
\par
 Studies on large supercells to achieve highly accurate evaluation of energies are usually limited to small systems (a few hundred atoms) and a subset of possible structures because of the high computational cost, meaning it is difficult to be sure of fully exploring the phase space. Simpler, atomistic model studies are able to explore the phase space more effectively provided suitable atom-atom pair interactions are available.  For multicomponent alloys, interactions are often assumed to be either pseudobinary, i.e. taken from analysis of related binary alloys, or limited to nearest neighbour interactions only, or both. There is scope for a rigorous procedure to obtain and test parameters for use in atomistic modelling.
\par
Here we address these topics. Our calculations represent the first computational implementation of an {\it ab initio} linear response theory for multicomponent alloys in a concentration wave formalism, which obtains the short-range order parameters directly and includes fully the effects from the rearrangement of the charge that occurs in response to the concentration wave modulation~\cite{staunton_compositional_1994,khan_statistical_2016} as well as the changes to the electronic energies. This is based on calculation of correlation functions and atomic short-range order {\it ab initio} from a Gibbs free energy. The linear response theory allows for extraction of atom-atom interchange parameters in a true multicomponent setting up to an arbitrary number of neighbour distances. 
\par
To demonstrate the effectiveness of our approach, we choose to study the so-called Cantor alloy---NiCoFeMnCr---and its derivatives as classified by Wu~\cite{wu_recovery_2014}, collectively referred to as the Cantor-Wu alloys~\cite{billington_bulk_2020}. This is a family of two, three, four, and five equiatomic component alloys forming FCC solid solutions, which includes NiCo, NiFe, NiCoCr, NiCoMn, NiCoMnCr, and NiCoFeCr. These materials are of interest for applications because many possess exceptional mechanical properties including hardness; resistance to wear, irradiation and fracture; and tensile strength~\cite{tsai_physical_2013, gludovatz_fracture-resistant_2014, zhang_microstructures_2014, pickering_high-entropy_2016}. This makes them well suited for a variety of next-generation engineering applications~\cite{senkov_development_2016, barron_towards_2020}. These systems also pose interesting fundamental physical questions because they are composed of mixtures of $3d$ transition metals, and it is known that in these materials magnetism affects both crystal structure and bulk mechanical properties.
We study an indicative series of alloys of increasing configurational entropy: NiCo, NiCoCr, NiCoFeCr, and NiCoFeMnCr, all of which have been shown to form single-phase FCC solid solutions down to low temperatures~\cite{wu_recovery_2014}. Previous works have suggested that the NiCoCr system is poorly modelled as pseudobinary~\cite{pei_statistics_2020}, so, in addition to NiCo, we also consider fictitious FCC NiCr and CoCr binary subsystems to better understand why this approximation fails. We show that, indeed, for the ternary system and higher, the approximations that the system can be modelled as pseudobinary or that interactions are short range and limited to nearest neighbour distance only are not well-founded. We elucidate the nature of short-range order, predict order/disorder transition temperatures, and also discuss the nature of the low-temperature ground state for these materials.
\par
This paper is laid out in the following way. First, in Sec.~\ref{sec:theory}, we provide an outline of the linear response theory used to produce our results ~\cite{khan_statistical_2016}. We also discuss the connection between the {\it ab initio} theory and atomistic modelling parameters. Then, in Sec.~\ref{sec:results}, we give results of electronic density of states calculations for the solid solutions, linear response analysis, and atomistic modelling of the phase stability for all four of the Cantor-Wu alloys. Rather than just providing predictions for the nature of compositional order in these materials, we also use details of the materials' electronic structure to elucidate its origins. Finally, in Sec.~\ref{sec:conclusions}, we summarise our results and give an outlook on further work.

\section{Theory}
\label{sec:theory}

This work represents the first results from a complete computational implementation of a theory of phase-stability of multicomponent alloys via electronic structure calculations using the Korringa Kohn Rostoker (KKR) method and the coherent potential approximation (CPA) developed in an earlier work~\cite{khan_statistical_2016}. This theory in turn is a generalisation of the $S^{(2)}$ direct correlation function theory for binary alloys~\cite{gyorffy_concentration_1983, staunton_compositional_1994}. Here we briefly outline the details of the theory and how it can be applied to the Cantor-Wu family of alloys.
\par
Compositional order in a substitutional alloy with fixed underlying lattice is described by set of variables referred to as site-wise occupation numbers. For a given configuration these are written $\{ \xi_{i \alpha} \}$, where $\xi_{i \alpha} = 1$ if site $i$ is occupied by an atom of species $\alpha$ and $\xi_{i \alpha} = 0$ otherwise. We note the physical constraint that $\sum_\alpha \xi_{i\alpha} = 1$, i.e. each site is occupied by one atom. The total concentration of a given species is given by $c_\alpha = \frac{1}{N} \sum_i \xi_{i\alpha}$, where N is the total number of lattice sites

\subsection{Alloy Grand Potential}
From statistical mechanics, the grand potential for an alloy is written as
\begin{equation}
\begin{aligned}
    \Omega &=  -\frac{1}{\beta} \ln \mathcal{Z} \\
    &= -\frac{1}{\beta} \ln \sum_{\{\xi_{i\alpha}\}} e^{-\beta ( \Omega_\text{el}[\{\xi_{i\alpha}\}] + \sum_{i\alpha} \nu_{i\alpha} \xi_{i\alpha})},
    \label{eq:grand_potential}
\end{aligned}
\end{equation}
so that the probability for a particular configuration $P\{\xi_{i\alpha}\}$ is given in terms of $\Omega_\text{el}[\{\xi_{i\alpha}\}]$ which represents the energy associated with the glue of interacting electrons moving in the static field of nuclei arranged on the lattice whose configuration is specified by $\{\xi_{i\alpha}\}$. $\Omega_\text{el}[\{\xi_{i\alpha}\}]$ is what would be computed in a density functional theory (DFT) calculation and consequently requires minimization with respect to charge and magnetization densities 
\begin{equation}
\frac{\delta \Omega_\text{el}\left[ \{\xi_{i\alpha}\} \right]}{\delta \rho({\bf r})}=0, \hspace{1cm} \frac{\delta \Omega_\text{el}\left[\{\xi_{i\alpha}\}\right]}{\delta {\bf M}({\bf r})}=0.
\label{eq:dft-min}
\end{equation}
The densities $\rho({\bf r};\{\xi_{i\alpha}\})$, ${\bf M}({\bf r};\{\xi_{i\alpha}\})$ are thus configuration-dependent.
The $\nu_{i\alpha}$ are site-wise chemical potentials, which are given both species and site indices to reflect how they are used in the linear response theory. In the homogeneous limit they are Lagrange parameters fixing the overall concentration of each species. At this point we could in principle choose some approximate parameterisation of $\Omega_\text{el}$---for example, in terms of pairwise interactions. However, for the purposes of formulating correlation functions rigorously we choose to keep this term in all its complexity. The two-point correlation function which describes atomic short-range order (ASRO), is given by
\begin{equation}
    \Psi_{i\alpha;j\alpha'} = \langle \xi_{i\alpha} \xi_{j\alpha'} \rangle - \langle \xi_{i\alpha} \rangle \langle \xi_{j\alpha'} \rangle,
\end{equation}
where angled brackets denote an ensemble average. This can be obtained directly from the free energy $\Omega$ via 
\begin{equation}
\frac{\partial^2 \Omega}{\partial \nu_{i\alpha} \partial \nu_{j\alpha'}} = -\beta \Psi_{i\alpha;j\alpha'}.
\end{equation}
\par
For the purpose of developing a Landau-type theory of compositional order, we seek to perform a series expansion about an uncorrelated reference state where $P\{\xi_{i\alpha}\}= \prod_i P_i(\xi_{i\alpha})$. That is, a state in which the distribution is generated by a mean-field Hamiltonian, $H_0 = \sum_{i \alpha} \mu_{i \alpha} \xi_{i\alpha}$, where the $\mu_{i\alpha}$ are to be determined. Starting from equation~\ref{eq:grand_potential},  it is possible to Taylor expand and write
\begin{align}
    \Omega &\approx -\frac{1}{\beta} \ln \sum_{\{\xi_{i\alpha}\}} (1 -\beta ( \Omega_\text{el} - H_0))e^{ - \beta( H_0 + \sum_{i\alpha} \nu_{i\alpha} \xi_{i\alpha})} \nonumber \\
    &\approx \Omega_0 + \langle \Omega_\text{el} - H_0 \rangle_0 \equiv \Omega^{(1)}.
\end{align}
By the Gibbs-Bogoliubov-Feynman inequality, $\Omega^{(1)}$ is an upper bound on the true free energy~\cite{feynman_statistical_1998}. The angle brackets $\langle \cdot \rangle_0$ denote an average taken with respect to the ensemble generated by the mean-field Hamiltonian $H_0$. This Hamiltonian naturally defines a set of average site occupancies, $\langle \xi_{i \alpha} \rangle_0 = \sum_{\alpha} \xi_{i \alpha} P_i(\xi_{i \alpha})=\Bar{c}_{i\alpha}$. The optimal uncorrelated reference system, i.e. specified by the best $\mu_{i \alpha}$ values, is found by minimizing $\frac{\partial \Omega^{(1)}}{\partial \Bar{c}_{i\alpha}} = 0$. From this expression, we write our approximation to the true free energy as
\begin{equation}
\begin{aligned}
    \Omega^{(1)}[\{\nu_{i\alpha}\}, \{\Bar{c}_{i\alpha}\}] = &-\frac{1}{\beta} \sum_{i\alpha} \Bar{c}_{i\alpha} \ln \Bar{c}_{i\alpha} - \sum_{i\alpha} \nu_{i\alpha} \Bar{c}_{i\alpha} \\ &+ \langle \Omega_\text{el} \rangle_0 [\{\Bar{c}_{i\alpha}\}],
\end{aligned}
\end{equation}
where the first term is the configurational entropy, sometimes also referred to as the entropy of mixing~\cite{chaikin_principles_1995}. The final term denotes the average value of the electronic and nuclear contribution to the free energy, where the average is taken with respect to the ensemble generated by the mean-field Hamiltonian, $ \prod_i  \sum_{\alpha} P_i(\xi_{i \alpha}) \Omega_\text{el}\{\xi_{i\alpha}\} =\langle \Omega_\text{el} \rangle_0 [\{\Bar{c}_{i\alpha}\}]$.

\subsection{Landau Theory}
\begin{figure*}
\centering
\includegraphics[width=0.3\textwidth]{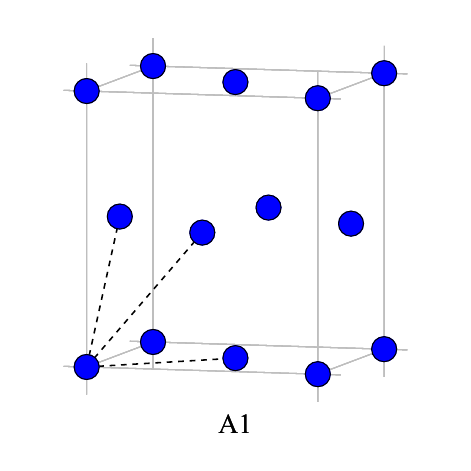}
\includegraphics[width=0.3\textwidth]{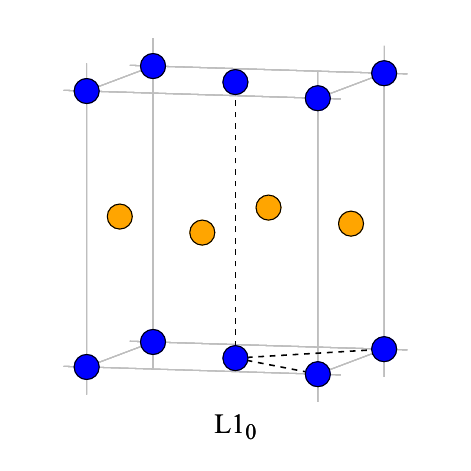}
\includegraphics[width=0.3\textwidth]{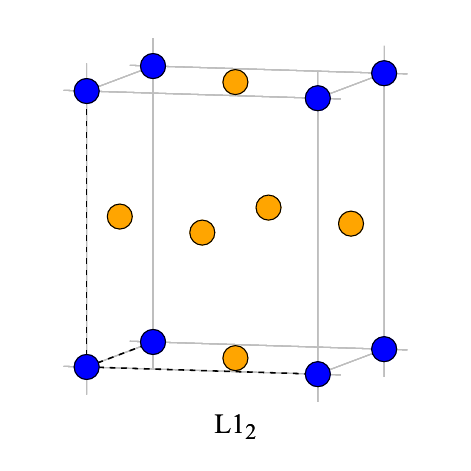}
\caption{Examples of ordered structures based on the FCC lattice which can be described by concentration waves. A1 represents a disordered alloy. L1\textsubscript{0} can be described by modes $\mathbf{k}=\{(0,0,1), (0,0,-1)\}$, while L1\textsubscript{2} is described by a combination of all six $\mathbf{k}$-vectors, $\mathbf{k}=\{0,0,1\}$.}
\label{fig:concentration_waves}
\end{figure*}
The Landau theory description is based on an expansion of the free energy about a homogeneous (high $T$) reference state. Writing $\Bar{c}_{i\alpha} = c_\alpha + \Delta \Bar{c}_{i\alpha}$, we are able to expand the free energy $\Omega^{(1)}$ as
\begin{equation}
\begin{aligned}
    \Omega^{(1)}(\{\Bar{c}_{i\alpha}\}) &= \Omega^{(1)}(\{c_{\alpha}\}) + \sum_{i\alpha} \left. \frac{\partial \Omega^{(1)}}{\partial \Bar{c}_{i\alpha}} \right|_{\{c_{\alpha}\}} \Delta \Bar{c}_{i\alpha} \\ 
    + \frac{1}{2} &\sum_{i\alpha; j\alpha'} \left. \frac{\partial^2 \Omega^{(1)}}{\partial \Bar{c}_{i\alpha} \partial \Bar{c}_{j\alpha'}} \right|_{\{c_{\alpha}\}} \Delta \Bar{c}_{i\alpha}\Delta \Bar{c}_{j\alpha'} + \dots
\label{eq:landau}   
\end{aligned}
\end{equation}
The first order term vanishes by considering the various symmetries of the homogeneous reference state, and we are left with the second order terms and higher. When applying an expansion such as this to the free energy of our uncorrelated state, the key quantity to evaluate comes from the second order concentration derivative of the ensemble averaged DFT energy $\langle \Omega_\text{el} \rangle_0 [\{\Bar{c}_{i\alpha}\}]$ namely $-\frac{\partial^2 \langle \Omega_\text{el} \rangle_0}{\partial \Bar{c}_{i\alpha} \partial \Bar{c}_{j\alpha'}} \equiv S^{(2)}_{i\alpha;j\alpha'}$ which is the the direct correlation function~\cite{gyorffy_concentration_1983, khan_statistical_2016}. The second order term of the full free energy, Eq.~\ref{eq:grand_potential} is related to the ASRO, and now we estimate this by $\Bar{\Psi}_{i\alpha;j\alpha'} = \beta^{-1} \frac{\partial \Bar{c}_{i\alpha}}{\partial \nu_{j\alpha'}}$. It should be emphasised that if the electronic and nuclear energy term in Eq.~\ref{eq:grand_potential} for the free energy,$\Omega_\text{el}\left[ \{\xi_{i\alpha}\}\right]$, could be written purely in terms of a sum of pairwise interactions, i.e. $ -\frac{1}{2} \sum_{i,\alpha;j\alpha'} V_{i,\alpha;j\alpha'} \xi_{i\alpha} \xi_{j\alpha'}$, then the $S^{(2)}$s would coincide with the $V$ parameters.
\par
Dropping higher order terms, the change in free energy due to a fluctuation $\{ \Delta \Bar{c}_{i\alpha} \}$ is written
\begin{equation}
\begin{aligned}
    \delta \Omega^{(1)} &= \frac{1}{2} \sum_{i,j} \sum_{\alpha, \alpha'} \Delta \Bar{c}_{i\alpha} [\beta^{-1} \Bar{\Psi}^{-1}_{i\alpha;j\alpha'}] \Delta \Bar{c}_{j\alpha'} \\
    &= \frac{1}{2} \sum_{i,j} \sum_{\alpha, \alpha'} \Delta \Bar{c}_{i\alpha} [\beta^{-1} \, C_{\alpha,\alpha'}^{-1} - S^{(2)}_{i\alpha, j\alpha'}] \Delta \Bar{c}_{j\alpha'},
    \label{eq:chemical_stability_real}
\end{aligned}
\end{equation}
where $C_{\alpha \alpha'}^{-1} = \frac{\delta_{\alpha \alpha'}}{c_\alpha}$ is associated with the entropic contributions. To make progress, we take advantage of the translational symmetry of the homogeneous reference state and Fourier transform. Using the language of concentration waves, we have that the change in free energy from a fluctuation $\delta \Bar{c}_\alpha({\bf k})$ is given by
\begin{equation}
\begin{aligned}
    \delta \Omega^{(1)} &= \frac{1}{2} \sum_{\bf k} \sum_{\alpha, \alpha'} \Delta \Bar{c}_\alpha({\bf k}) [\beta^{-1} \Bar{\Psi}^{-1}_{\alpha\alpha'}({\bf k})] \Delta \Bar{c}_{\alpha'}({\bf k}) \\
    &= \frac{1}{2} \sum_{\bf k} \sum_{\alpha, \alpha'} \Delta \Bar{c}_\alpha({\bf k}) [\beta^{-1} C^{-1}_{\alpha \alpha'} -S^{(2)}_{\alpha \alpha'}({\bf k})] \Delta \Bar{c}_{\alpha'}({\bf k}).
\end{aligned}
\label{eq:chemical_stability}
\end{equation}
We refer to the matrix in the square brackets as the chemical stability matrix, although it is clear from Eq.~\ref{eq:chemical_stability_real} and Eq.~\ref{eq:chemical_stability} that it can equally well be interpreted as the inverse of the two-point correlation function, our SRO parameter. This is intuitive; it implies that the cost of a fluctuation along a particular mode is inversely proportional to the SRO parameter, which makes sense because the SRO parameter measures the tendency of atoms to cluster or kept apart.
\par
It is worth taking a brief aside to give a few examples of how structures can be classified by particular concentration waves, as proposed by Khachaturyan and others~\cite{khachaturyan_ordering_1978}. For the FCC lattice, three illustrative examples for two-component alloys are A1, L1\textsubscript{0}, and L1\textsubscript{2}, illustrated in Fig.~\ref{fig:concentration_waves}. A1 represents a disordered, high temperature state in which no long-range order is present and lattice sites are occupied by each species with equal probability. For a binary system, the L1\textsubscript{0} structure, commonly referred to as CuAu can be described by a superposition of two concentration waves with polarisation $\delta c_\alpha = (1,-1)$ in concentration space and modes $k_0 = (0,0,1), (0, 0, -1)$. The L1\textsubscript{2} structure, in which atoms of one species occupy the corners of the cube and which is commonly referred to as Cu$_3$Au, is described again by a concentration wave with polarisation $\delta c_\alpha = (1,-1)$ but this time six equivalent $\mathbf{k}$-vectors, $\mathbf{k}=\{0,0,1\}$. It should be stressed that for a true L1\textsubscript{0} structure we require equal concentrations of species $A$ and $B$, while the L1\textsubscript{2} structure requires a composition of $A_3B$. However, it is possible for non-stoichiometric compositions to exhibit L1\textsubscript{0}- and L1\textsubscript{2}-like order, including in multi-component systems where, for example, one chemical species might occupy one sub-lattice preferentially with the other species disordered on other sites, as will be seen in Sec.~\ref{sec:results}.
\par
Once we have the chemical stability matrix, shown in Eq.~\ref{eq:chemical_stability}, in reciprocal space, there are two ways by which to proceed. The first is to look directly at the eigenvalues and eigenvectors of the chemical stability matrix in reciprocal space and investigate the ASRO. Once the lowest eigenvalue of this matrix passes through zero at mode ${\bf k_{\text{{us}}}}$ at some temperature $T_0$, we infer that the homogeneous state is unstable to a fluctuation along that mode. The associated eigenvalue and eigenvectors inform us of the dominant correlations along that mode above $T_0$. These eigenvalues and eigenvectors require careful interpretation to ensure that overall concentrations of each species are conserved by a given fluctuation. In particular, the sum rule $\sum_\alpha c_{i\alpha} = 1$ must be obeyed, which means that for $s$ species there are $s-1$ independent variables on each site~\cite{singh_atomic_2015}. For an $s \times s$ chemical stability matrix, we perform an orthogonal transformation to isolate this unphysical degree of freedom, leading to a matrix with $s-1$ physically meaningful eigenvalues~\cite{khan_statistical_2016}.
\par
It should be emphasised that the response theory and eigenvalue/eigenvector stability analysis, which is valid above any order/disorder transition temperature $T_0$ and describes ASRO in the solid solution phase, does not make any presupposition about the dependence of $\Omega_\text{el}$ on the atomic site occupations $\{ \xi_{i\alpha}\}$. The $S^{(2)}_{i\alpha;j\alpha'}$s are rigorously related to the correlation functions and the leading coefficients in the Landau expansion of the free energy about the homogeneously disordered phase.

\subsection{The Direct Correlation Functions {\bf \emph{ab initio}}}
We evaluate the key quantities, the direct correlation functions $S^{(2)}_{i\alpha;j\alpha'}$, following the scheme explained in detail for multicomponent alloys in Ref.~\citenum{khan_statistical_2016}.  We start with an approximation of the internal energy of the system of interacting electrons and nuclei averaged over configurations, $\langle \Omega_\text{el} \rangle_0 [\{\Bar{c}_{i\alpha}\}]$, within the tenets of DFT. This energy depends on the concentrations $\{ \Bar{c}_{i\alpha} \}$ and charge and magnetization densities, $\{ \Bar{\rho}_{i\alpha}({\bf r}_i) \}$, $\{ \Bar{{\bf M}}_{i\alpha}({\bf r}_i) \}$. $\Bar{\rho}_{i\alpha}({\bf r}_i)$ is the charge density $\rho({\bf r}_i)$ in the space around site $i$ averaged over all configurations which have the atomic species $\alpha$ occupying site $i$.  We assume that the alloys are in or quenched from their paramagnetic states which we describe within the Disordered Local Moment (DLM) picture~\cite{gyorffy_first-principles_1985}.

The DFT step of Eq.~\ref{eq:dft-min} requires $\langle \Omega_\text{el} \rangle_0 [\{\Bar{c}_{i\alpha}\}] = \Bar{\Omega}[\{\Bar{c}_{i\alpha}\},\{ \Bar{\rho}_{i\alpha}({\bf r}_i) \},\{ \Bar{{\bf M}}_{i\alpha}({\bf r}_i) \}]  $ to be minimized with respect to partially averaged charge and magnetization densities:
\begin{equation}
\frac{\delta \Bar{\Omega}[\{\Bar{c}_{i\alpha}\}}{\delta \Bar{\rho}_{j\gamma}({\bf r}_j)}=0, \hspace{1cm} \frac{\delta \Bar{\Omega}[\{\Bar{c}_{i\alpha}\}}{\delta {\Bar{\bf M}}_{j\gamma}({\bf r}_j)} =0.
\label{eq:DFTapprox}
\end{equation}
The charge and magnetization densities associated with any given site, $j$, occupied by an atom of species $\gamma$, are consequently dependent on the concentrations on all other sites, i.e. $\Bar{\rho}_{j\gamma}({\bf r}_j) = \Bar{\rho}_{j\gamma}({\bf r}_j;\{\Bar{c}_{i\alpha}\}, i \, \rm{excluding} \, j)$. From Eq.~\ref{eq:landau} we expand $\Bar{\Omega}[\{\Bar{c}_{i\alpha}\}]$ around the homogeneously disordered state where $\{ \Bar{c}_{i\alpha}= c_{\alpha}\}$, $\{ \Bar{\rho}_{j\alpha}({\bf r}_j) = \Bar{\rho}_{\alpha}({\bf r}_j) \}$ \textit{etc.} For a set of small concentration fluctuations $\{ \Delta c_{i\alpha} \}$, where the overall concentrations are unchanged,
\begin{widetext}
\begin{equation}
\begin{aligned}
\Bar{\Omega}[\{c_{i\alpha}\}] & = \Bar{\Omega}_0[\{\Bar{c}_{\alpha}\}] + \sum_{i\alpha} \frac{\partial \Bar{\Omega}}{\partial \Bar{c}_{i\alpha}}  \Delta c_{i\alpha} + \sum_{i\alpha} \sum_{k\gamma} \int d {\bf r}_k \frac{\delta \Bar{\Omega}}{\delta \Bar{\rho}_{k\gamma}({\bf r}_k)} 
\frac{\partial \Bar{\rho}_{k\gamma}({\bf r}_k)}{\partial \Bar{c}_{i\alpha}}
\Delta c_{i\alpha} 
\\&+ \frac{1}{2}\sum_{i\alpha} \sum_{j\alpha'} 
[\frac{\partial^2 \Bar{\Omega}}{\partial \Bar{c}_{i\alpha}\partial \Bar{c}_{j\alpha'}}  + \sum_{k\gamma} \int d {\bf r}_k  \frac{\partial}{\partial \Bar{c}_{k\gamma} } \large(\frac{\delta \Bar{\Omega}}{\delta \Bar{\rho}_{k\gamma}({\bf r}_k)}\large) \frac{\partial \Bar{\rho}_{k\gamma}({\bf r}_k)}{\partial \Bar{c}_{j\alpha'}}]  \Delta c_{i\alpha} \, \Delta c_{j\alpha'}.
\end{aligned}
\label{eq:expand}
\end{equation}
\end{widetext}
The second and third terms of Eq.~\ref{eq:expand} vanish owing to fixed overall concentrations, $\sum_i \Delta c_{i\alpha} =0$, and the DFT charge stationarity conditions, Eq.~\ref{eq:DFTapprox} respectively. The terms in square brackets (4th and 5th terms) constitute the direct correlation function, $S^{(2)}_{i\alpha, j\alpha'}$~\cite{khan_statistical_2016}.  It includes effects not only from how the single particle energies contribution to the energy $\Bar{\Omega}[\{\Bar{c}_{i\alpha}\}]$ depends on the configurational configuration (the `band-energy' component to $S^{(2)}_{i\alpha, j\alpha'}$~\cite{singh_atomic_2015, singh_ta-nb-mo-w_2018}) but also on how the electronic charge rearranges with fluctuations in composition. This latter piece depends on the charge responses $\frac{\partial \Bar{\rho}_{k\gamma}({\bf r}_k)}{\partial \Bar{c}_{j\alpha'}}$. In our calculations, an Onsager reaction field ensures that $\frac{\partial \Bar{\rho}_{j\gamma}({\bf r}_j)}{\partial \Bar{c}_{j\alpha'}} =0$ as required by the definition of $\Bar{\rho}_{j\gamma}({\bf r}_j = \Bar{\rho}_{j\gamma}({\bf r}_j;\{\Bar{c}_{i\alpha}\}, i \, \rm{excluding} \, j)$. In principle the response of the magnetization densities should also be included but in this work on paramagnetic alloys they are small and we neglect them.

Often it is a good approximation for many metallic alloys to assume that the `band energy' contribution to the $S^{(2)}_{i\alpha;j\alpha'}$'s is enough to describe adequately compositional ordering and atomic short range order but we find that this approximation falls short for the multi-component Cantor-Wu high entropy alloys of this study - the effects from how the electronic charge rearranges, when changes to the ensemble of configurations are made, turn out to be significant.

\subsection{Atomistic Modelling}
\label{sec:theory-mc}
To further investigate the phase behaviour of the alloys, we map the energetics onto a simpler, pairwise Hamiltonian of the form
\begin{equation}
    H = -\frac{1}{2}\sum_{i \alpha; j\alpha'} V_{i\alpha; j\alpha'} \xi_{i \alpha} \xi_{j \alpha'} + \sum_{i\alpha} \nu_{i\alpha} \xi_{i\alpha}, 
    \label{eq:b-w}
\end{equation}
as originally proposed by Bragg and Williams~\cite{bragg_effect_1934, bragg_effect_1935}. For such a model the $V_{i\alpha; j\alpha'}$s are equivalent to the $S^{(2)}_{i\alpha;j\alpha'}$s, and when this quantity is calculated in reciprocal space the $V_{i\alpha; j\alpha'}$ are recovered by fitting to a real-space interaction.  It should be emphasised that the earlier instability analysis is only rigorous for second-order transitions, but these $V$s can be used to infer transitions which are first-order.  By identifying the $S^{(2)}_{i\alpha;j\alpha'}$s  with this mapping, we have atom-atom interchange parameters that can be used for modelling at any temperature. This step assumes that these parameters are the same in the high-$T$, homogeneous limit as for low-$T$ states with order developing. The procedure by which these pairwise interactions are obtained, via analysis of the free energy cost of compositional fluctuations around the disordered phase, makes them an unbiased best choice, being unrelated to fits to energies of specific configurations. 
\par
Because we relate $V_{i\alpha; j\alpha'}$ directly to our $S^{(2)}_{\alpha \alpha'}(\mathbf{k})$s, evaluated for many $\mathbf{k}$ values, we are able to define these quantities unambiguously and also determine their extent beyond nearest neighbour distance. (In practice we observe the convergence of our fit and use this to decide how many real space shells the interaction should be fitted to.) Many studies, however, approximate $V_{i\alpha; j\alpha'}$ either as restricted to nearest neighbour distance or as pseudobinary, or both~\cite{santodonato_predictive_2018, troparevsky_criteria_2015}. The `pseudobinary' approximation, that the interaction between species $A$ and species $B$ is the same in a multicomponent system as for the binary, needs careful scrutiny. The presence of an additional species will alter the underly electronic structure of the system and potentially affect the interactions between species as a consequence. Moreover, the $A$-$B$ binary may not have the same underlying lattice as the multicomponent system. In this work, our calculations enable us to model interactions in a true multicomponent setting.
\par
With the extraction of pairwise interactions from our Landau-type theory, we can study an atomistic model to explore the phase space. To that end, we use the Metropolis Monte-Carlo algorithm with the so-called Kawasaki dynamics to conserve overall concentrations of each species~\cite{landau_guide_2014}. This method has been used with success to study the physics of alloy formation~\cite{binder_monte_1981, santodonato_predictive_2018}. A lattice model is well-suited to this family of alloys because it is known from both experiment and theory that lattice distortions in these systems are small~\cite{oh_lattice_2016}.

\begin{figure*}[ht]
\centering
\includegraphics[width=\textwidth]{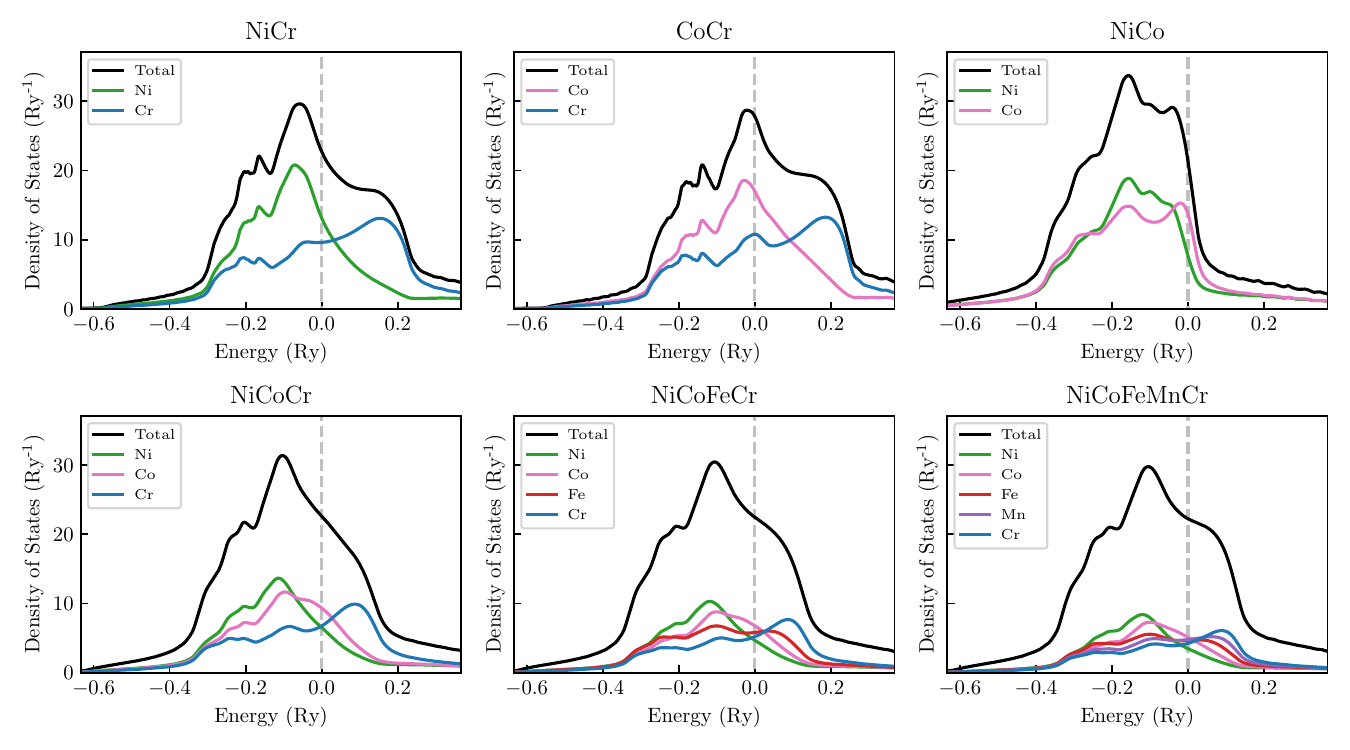}
\caption{Comparison of the species-resolved density of states of the systems considered in this work. We use the CPA to average over disorder, and the DLM picture to reflect that we expect these systems to be in a paramagnetic state when prepared. This results in fine detail being smeared out. The most notable feature is how the addition of Cr completely changes the DoS curves for both Ni and Co when compared to the binary NiCo, indicating that the Ni-Co interaction is likely to be poorly approximated as pseudobinary in multicomponent systems.}
\label{fig:dos_comparison}
\end{figure*}

The system configuration is specified by the occupation numbers $\{\xi_{i\alpha}\}$, with internal energy given by the Hamiltonian as in Eq.~\ref{eq:b-w}. The setup is perhaps best thought of as an Ising-like model, but one in which there is a vector of occupation numbers on each site, rather than just a single spin. The site occupancies are initialised at random, with the only restrictions being the overall concentration of each species. The algorithm then proceeds as follows. A pair of lattice sites (not necessarily nearest neighbours) are selected at random, and the change in energy $\Delta H$ from swapping the site occupancies is computed. If the change in energy is negative the move is accepted unconditionally, while if the change is positive the swap is accepted with probability $e^{-\beta \Delta H}$.
Care must be taken to ensure that equilibrium is achieved at a given temperature, and that any results are well-converged with respect to system size. Our implementation applies periodic boundary conditions in all three directions. A measure of the configurational contribution to the specific heat capacity (SHC) can be obtained via the fluctuation-dissipation theorem~\cite{allen_computer_2017}. In equilibrium, an estimation of the specific heat is given by
\begin{equation}
    C = \frac{1}{k_b T^2} \left( \langle E^2 \rangle - \langle E \rangle^2 \right),
\end{equation}
and it is this which we calculate to obtain our SHC curves. We infer a phase transition when a peak in the SHC curve is observed.
\par
To quantify short-range order (SRO) in our simulations, we take an equilibrated configuration and calculate for calculate the pairwise probabilities of finding atoms at nearest and next-nearest neighbour distance. For example, in a ternary $ABC$ alloy we calculate the fraction of $A$-$A$, $A$-$B$, $A$-$C$, $B$-$B$, $B$-$C$, and $C$-$C$ pairs at nearest and next-nearest neighbour distance. In a sample in which no SRO is present, we expect all of these to take the value $\frac{1}{3}$, but when SRO emerges they will move away from these values.

\section{Results and Discussion}
\label{sec:results}

\begin{figure*}[ht]
\centering
\includegraphics[width=\textwidth]{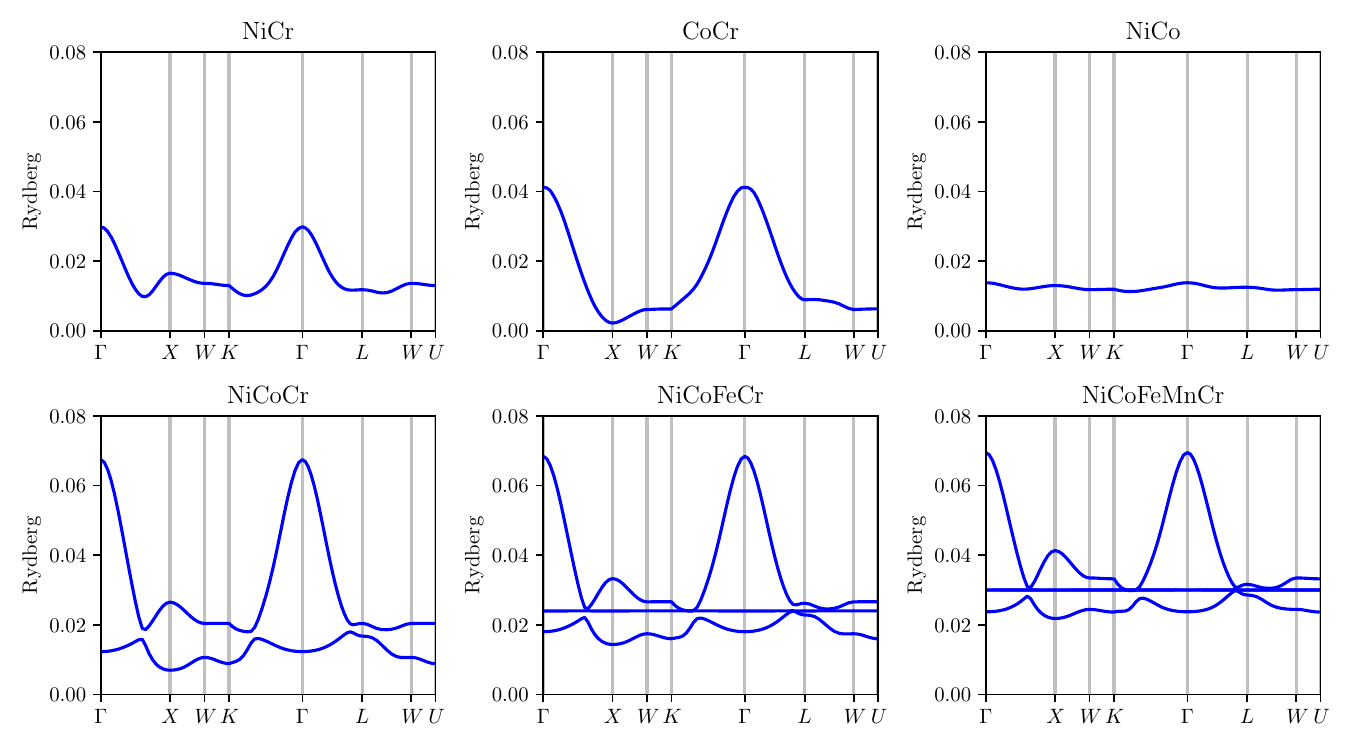}
\caption{Eigenvalues of the chemical stability matrix along symmetry lines around the Brillouin zone. All are computed at a temperature of 1000K, above any predicted ordering temperature. For an $s$-species alloy there are $s-1$ physically meaningful eigenvalues due to the conservation of overall concentrations. In line with the trends shown in the density of states plots, the contrast between the binary alloys and those with three, four, or five components can clearly be seen. It should also be noted that the addition of Fe and Mn does not significantly alter the shape of the two modes present in NiCoCr. Rather, Fe and Mn introduce two essentially flat modes which, in concentration space, are polarised along the Fe and Mn directions indicating that these elements dilute the other interactions in the system. The upward trend in the lowest-lying eigenvalue from the 3-component through to the 5-component system is a result of the increasing contribution from the configurational entropy stabilising the solid solution.}
\label{fig:eigenvalues}
\end{figure*}

\subsection{Density of States}
The first stage of our workflow is to generate the self-consistent one-electron potentials and densities of DFT for the disordered multi-component alloys in their paramagnetic states since these materials only exhibit magnetic order at low temperatures. Earlier studies of magnetism in these systems have noted that all have low Curie temperatures, usually below room temperature~\cite{billington_bulk_2020, sales_quantum_2016, kao_electrical_2011, schneeweiss_magnetic_2017} and well below the temperatures at which the alloys are prepared. This stage is carried out using the HUTSEPOT code~\cite{hoffmann_magnetic_2020}, although in principle any KKR-CPA code is suitable. We perform scalar-relativistic calculations within the atomic sphere approximation (ASA)~\cite{stocks_complete_1978} with an angular momentum cutoff of $l_\text{max} = 3$ for basis set expansions, a $20\times20\times20$ Monkhorst-Pack grid~\cite{monkhorst_special_1976} for integrals over the Brillouin zone, and a 24 point semi-circular Gauss-Legendre grid in the complex plane to integrate over valence energies. We use the local density approximation (LDA) and the exchange-correlation functional is that of Perdew-Wang~\cite{perdew_accurate_1992}. Lattice parameters of 3.53, 3.56, 3.57 and 3.59\AA \, are used for NiCo, NiCoCr, NiCoFeCr, and NiCoFeMnCr, respectively, consistent with their experimental values~\cite{zhang_influence_2015, yin_yield_2020, cantor_microstructural_2004}. For the additional NiCr and CoCr binaries considered, we set the lattice parameter to 3.53\AA \, for consistency with the NiCo calculation. The paramagnetic states are described within the Disordered Local Moment (DLM) picture~\cite{gyorffy_first-principles_1985}. 

Fig.~\ref{fig:dos_comparison} shows a comparison of the total density of states for the four compounds considered in this work plus the additional NiCr and CoCr binaries, species-resolved. Although the NiCo system forms a single-phase FCC solid solution, neither of the other binaries are expected to  do so at this stoichiometry~\cite{nishizawa_coni_1983, nash_crni_1986, ishida_co-cr_1990}. In our calculations we have modelled them as a single phase solid solution on the FCC lattice to understand the differences between interactions when modelling interactions as purely binary as compared to our true multicomponent theory.

When considering the three binaries, it is clear that NiCo is most likely to form a stable solid solution down to low temperatures because of how similar the DoS curves for the two components are across the entire valence electron energy range. By contrast, when Cr is included its species-resolved DoS curve has a peak above the Fermi energy, quite far away from either the Ni or Co peaks. Perhaps the most striking feature of these plots is the contrast between the NiCo binary and NiCoCr, where the addition of Cr significantly alters the shape of both the Ni and Co DoS curves. We expect this to have a significant effect on the strength of Ni-Co correlations when comparing the binary to the three component system. Even before calculating atomic correlations, this provides tangible evidence that the ternary NiCoCr system is likely to be poorly approximated in pseudobinary terms, which has been suggested in earlier works~\cite{pei_statistics_2020}. The addition of Fe and Mn does little to alter the overall profile of either the total DoS curve or the individual Ni, Co, and Cr species-resolved curves for the quarternary NiCoFeCr or the quinary NiCoFeMnCr, suggesting that Fe and Mn will mix well with the other elements and stabilize the solid solution.

\subsection{Linear Response Analysis}

\begin{table*}[ht]
\begin{ruledtabular}
\begin{tabular}{llllllll}
System     & $k$-vector      & $\delta c_1$ & $\delta c_2$ & $\delta c_3$ & 
$\delta c_4$ & $\delta c_5$ & $T_\text{order}$ (K) \\ \hline
NiCr       & $(0, 0, 0.6)$ & 0.70711 &-0.70711 &&&
                            & 200                   \\ 
CoCr       & $(0, 0, 1)$ & 0.70711 &-0.70711 &&&
& 793                   \\ 
NiCo       & $(0, 0.6, 0.6)$ & 0.70711 &-0.70711 &&&
                            & 83                   \\ 
NiCoCr     & $(0, 0, 1)$     & -0.034613 &-0.68916 & 0.72378 & &                 & 606                  \\ 
NiCoFeCr   & $(0, 0, 1)$     & 0.013671&-0.68858&-0.048489&0.72340&          & 404                  \\ 
NiCoFeMnCr & $(0, 0, 1)$     & 0.033024&-0.68516&-0.081509&0.010666&0.72298 & 281                  \\
\end{tabular}
\end{ruledtabular}
\caption{Calculated transition temperatures for our considered systems, including the mode and eigenvector for which the instability occurs. We note that transition temperatures are computed using a mean field theory and so are expected to be overestimates of their true values. The three, four, and five component systems show a clear trend that increasing configurational entropy stabilises the solid solution and drives down the ordering temperature. The binaries evidence the differing strength and nature of interactions between the various elements.}
\label{tab:ordering}
\end{table*}

Starting from the self-consistent potentials and charge densities from the solid solution, we implement the linear response theory described in Sec.~\ref{sec:theory} to obtain the atomic short-range order {\it ab initio}. We use these results to construct the chemical stability matrix around the Brillouin zone. Plots of the eigenvalues of this matrix along various symmetry lines of the FCC lattice for all considered systems are shown in Fig.~\ref{fig:eigenvalues}. What can be seen clearly is that the addition of Fe and Mn add near-flat modes to the eigenvalue spectrum. In concentration wave space, these flat modes are polarised along the Fe and Mn `directions', which we associate with the fact that these elements interact weakly with the others and serve to dilute the Ni-Co-Cr interactions.

Table~\ref{tab:ordering} then shows computed ordering temperatures and the associated modes and eigenvectors for which the instability occurs. That is, we incrementally reduce the temperature and at each temperature search the first Brillouin zone for the lowest eigenvalue and its associated eigenvector. When this eigenvalue passes through zero we infer the transition.

The three binary systems illustrate the contrasting relative strength of the Ni-Co, Ni-Cr, and Co-Cr interactions when constrained to an FCC lattice. The predicted modes for NiCr and NiCo are consistent with existing theoretical and experimental literature on SRO in these systems~\cite{rahaman_first-principles_2014,hirabayashi_experimental_1969, schonfeld_short-range_1988, schweika_neutron-scattering_1988}. For NiCo, both the associated energies and the transition temperature are very small, consistent with our DoS plots and experimental literature~\cite{nishizawa_coni_1983}. The addition of Cr to form the NiCoCr ternary causes significant change in the eigenvalues and eigenvectors of the chemical stability matrix, again demonstrating the shortcomings of the pseudobinary approximation.

The modes for which the instability occurs for CoCr, NiCoCr, NiCoFeCr, and NiCoFeMnCr are all $\mathbf{k}_0=(0,0,1)$, with the dominant components of the eigenvector associated with Co and Cr. This is indicative of Co and Cr setting up long range order in an L1\textsubscript{0} structure. L1\textsubscript{0} indicates alternating layers of Co-rich, Cr-depleted and Co-depleted, Cr-rich.  It can be seen from the table that the components of the eigenvectors associated with Ni, Fe, and Mn are all much smaller than other ones, meaning that the correlations between these and other elements are not the dominant ones. For the four component NiCoFeCr system, our results agree with existing computational studies, which find again that correlations between Co and Cr are dominant~\cite{schonfeld_local_2019, niu_spin-driven_2015, fukushima_local_2017}.

When the ternary, quarternary, and quinary systems are taken in sequence, there is a clear trend that increasing configurational entropy drives down the transition temperature for these systems, reflected in the increase in the value of the lowest lying eigenvalue in Fig.~\ref{fig:eigenvalues}. This can be understood by considering the competition between the two terms in Eq.~\ref{eq:chemical_stability}. Fe and Mn interact weakly with the other elements in the system, so their addition makes little contribution to $S^{(2)}_{\alpha \alpha'}(\mathbf{k})$, but the additional element(s) increases the size of the diagonal terms in the matrix given by $C^{-1}_{\alpha \alpha'}$. This can be thought of as akin to the process of matrix regularisation. Put simply, the addition of weakly interacting elements makes a notable contribution to the configurational entropy while affecting the energetics of the system very little, thus driving down the ordering temperature.

Significantly, if we `turn off' the effect of charge rearrangement in our linear response calculation (keeping so-called `band energy' only terms) we erroneously find that nickel would phase segregate out of the multicomponent alloys at high temperatures.

\subsection{Fitting to Pairwise Interactions}
\begin{table}[b]
\begin{ruledtabular}
\begin{tabular}{lllll}
$V_{\alpha \beta}^{(1)}$& Ni     & Co     & Fe     & Cr     \\ \hline
Ni                      & -0.338 & 0.606  & 0.097  & -0.367 \\
Co                      & 0.606  & 0.656  & -0.049 & -1.213 \\ 
Fe                      & 0.097  & -0.049 & -0.019 & -0.029 \\ 
Cr                      & -0.367 & -1.213 & -0.029 &  1.609 \\ 
&&&& \\
$V_{\alpha \beta}^{(2)}$& Ni     & Co     & Fe     & Cr     \\ \hline
Ni                      &  0.316 & 0.058  & -0.061 & -0.313 \\
Co                      & 0.058  & 0.005  & -0.007 & -0.057 \\
Fe                      & -0.061 & -0.007 & -0.010 &  0.058 \\
Cr                      & -0.313 & -0.057 & 0.058  &  0.312 \\
&&&& \\
$V_{\alpha \beta}^{(3)}$& Ni     & Co     & Fe     & Cr     \\ \hline
Ni                      &  0.002 & 0.090  &  0.008 & -0.100 \\ 
Co                      & 0.090  & 0.053  & -0.009 & -0.135 \\ 
Fe                      &  0.008 & -0.009 & -0.002 &  0.003 \\ 
Cr                      & -0.100 & -0.135 & 0.003  &  0.232 \\ 
&&&& \\
$V_{\alpha \beta}^{(4)}$& Ni     & Co     & Fe     & Cr     \\ \hline
Ni                      &  0.255 & 0.006  & -0.032 & -0.229 \\
Co                      & 0.006  & -0.046 & -0.001 & 0.041 \\
Fe                      & -0.032 & -0.001 &  0.003 & 0.030 \\
Cr                      & -0.229 &  0.041 &  0.030 &  0.158
\end{tabular}
\end{ruledtabular}
\caption{Fitted interchange parameters up to fourth neighbour distance for NiCoFeCr. (Computed parameters for the other alloys considered are provided in the supplementary material.) We find no significant improvement in the fit is obtained by going beyond the fourth neighbour shell. All values are quoted in mRy.}
\label{tab:fit}
\end{table}

\begin{table}[t]
\begin{ruledtabular}
\begin{tabular}{llll}
$V_{\alpha \beta}^{(1)}$ & Ni     & Co     & \\ \hline
Ni                       & 0.023  & -0.023 & \\
Co                       & -0.023 & 0.023  & \\
                         &        &        & \\
$V_{\alpha \beta}^{(1)}$ & Ni     & Cr     & \\ \hline
Ni                       & 0.199  & -0.199 & \\ 
Cr                       & -0.199 & 0.199  & \\ 
                         &        &        & \\  
$V_{\alpha \beta}^{(1)}$ & Co     & Cr     & \\ \hline
Co                       & 1.026  & -1.026 & \\ 
Cr                       & -1.026 & 1.026  & \\ 
                         &        &        & \\ 
$V_{\alpha \beta}^{(1)}$ & Ni     & Co     & Cr     \\ \hline
Ni                       & -0.218 & 0.682  & -0.465 \\
Co                       & 0.682  & 0.672  & -1.351 \\
Cr                       & -0.465 & -1.351 & 1.813
\end{tabular}
\end{ruledtabular}
\caption{Comparison of computed pairwise interactions at nearest neighbour distance for the NiCoCr ternary and its three binary subsystems. All energies are quoted in mRy. It can be seen that the ternary interaction is not readily constructed from the three binaries; there are substantial changes in the sizes of interaction strengths, particularly for the Ni-Co and Ni-Cr interactions.}
\label{tab:pseudobinary}
\end{table}
A lattice Fourier transform of the direct correlation function, $S^{(2)}_{\alpha\alpha'}({\bf k})$, from the linear response analysis and calculation, yields real-space pairwise interactions suitable for use in atomistic modelling. We sample 56 unique $\mathbf{k}$ points distributed in the irreducible wedge of the first Brillouin zone, including along the special directions. We then use the method of least squares to fit to a function of the form
\begin{equation}
    S^{(2)}_{\alpha \alpha'}(\mathbf{k}) \approx \sum_{n=0}^N V^n_{\alpha \alpha'} \left( \sum_{\left\{ R_i \right\}_n} e^{i \mathbf{k} \cdot \mathbf{R}_i}\right),
\end{equation}
where $\left\{ R_i \right\}_n$ denotes the set of vectors pointing to all lattice sites on the $n$th neighbour shell, and $N$ denotes the maximum number of shells considered. The $V^n_{\alpha \alpha'}$ are the coefficients fitted.
\par
Complete fits for the four physically realisable FCC systems (NiCo, NiCoCr, NiCoFeCr, and NiCoFeMnCr) are provided in the supplementary material.  There are two key points to be made here. Firstly, for all systems considered, we find interactions to be quite long-ranged and the best balance between accuracy and computational cost is given by a fit to four neighbour shells in the FCC lattice. Table~\ref{tab:fit} shows a sample fitted interaction for one of the systems, NiCoFeCr. While the strongest interactions are over nearest-neighbour range, they are still significant at second and third nearest neighbour distance. The dominant interactions are Ni-Ni, Cr-Cr, Co-Cr, Co-Co, and Ni-Co, while Fe interacts very weakly with both itself and the other species present. This again demonstrates that Fe serves to dilute the system and stabilise the solid solution. Secondly, 
Table~\ref{tab:pseudobinary} compares interactions for the NiCoCr system with those for the three possible binary subsystems, NiCo, NiCr, and CoCr, and demonstrates the limitations of a pseudobinary interpretation of the interactions in the ternary alloy.
\subsection{Atomistic Modelling}
\begin{figure*}
\centering
\includegraphics[width=0.8\textwidth]{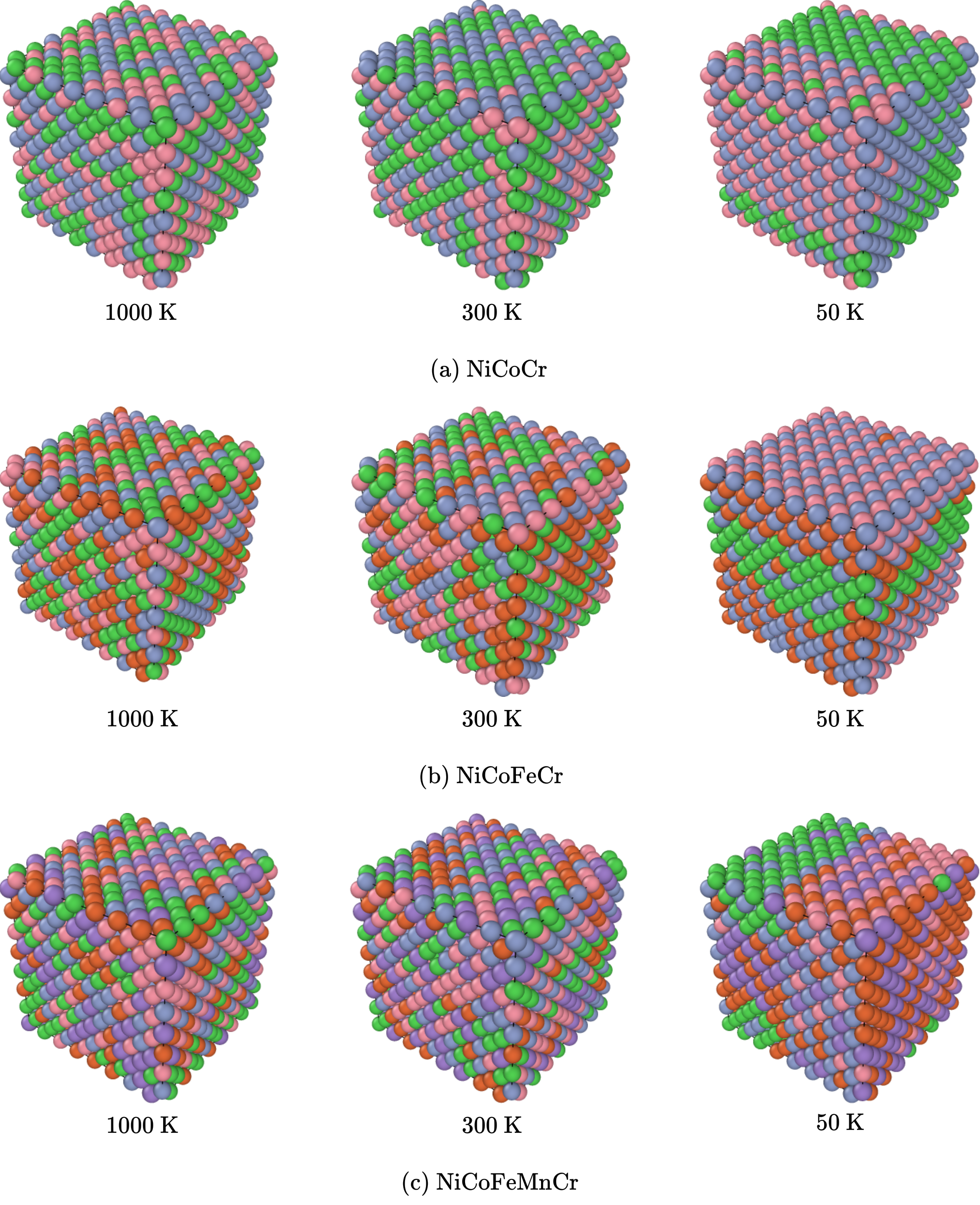}
\caption{Visualisations of indicative atomic configurations at a range of temperatures for NiCoCr, NiCoFeCr, and NiCoFeMnCr. Ni, Co, Fe, Mn, and Cr are indicated by green, pink, red, purple, and blue respectively. Each visualisation is of a sample configuration of 2048 atoms equilibrated at the denoted temperature. The visualised configurations, taken in combination with the SRO parameters visualised in Fig.~\ref{fig:sro} show that the low-temperature ground state of the system as multiphase, with some regions Ni-rich and some regions Ni-deficient, although this occurs at comparatively low temperatures and may not be experimentally accessible.}
\label{fig:visualisations}
\end{figure*}
\begin{figure*}
\centering
\includegraphics[width=0.8\textwidth]{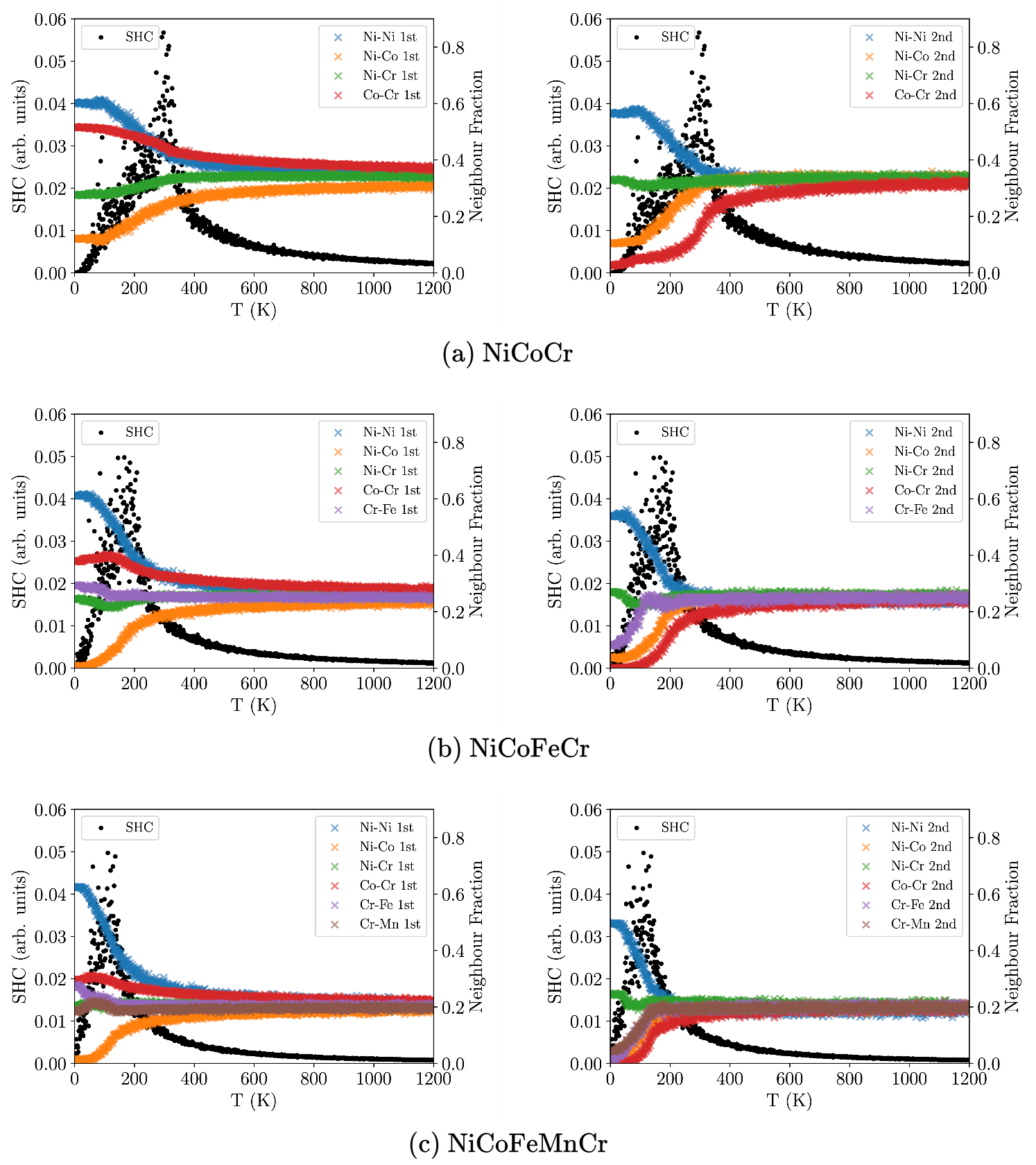}
\caption{Specific heat capacity curves (SHC) and short-range order (SRO) parameters for NiCoCr, NiCoFeCr, and NiCoFeMnCr. (Details of how SHC values and SRO parameters are calculated are given in Sec.~\ref{sec:theory-mc}.) In all three systems at low temperatures we find a predominance towards Ni-Ni and Co-Cr nearest neighbour pairs, and away from Ni-Co nearest neighbour and Co-Cr second nearest neighbour pairs. Very little ordering emerges between Fe, Mn and the other species present in either the quarternary or quinary, confirming that these elements interact weakly and stabilise the solid solution. Where multiple curves lie on top of each other at high temperatures it indicates little to no SRO.}
\label{fig:sro}
\end{figure*}
From our obtained pairwise interactions, we ran Monte Carlo (MC) simulations for NiCoCr, NiCoFeCr, and NiCoFeMnCr. All three systems consisted of 2048 atoms with periodic boundary conditions applied and were prepared in an initial, random configuration and annealed from 1200K in steps of 1K, with $10^3$ MC steps per atom at each temperature. The code used was developed in-house and follows the prescription described in Sec.~\ref{sec:theory-mc}.
\par
Figs.~\ref{fig:visualisations} and \ref{fig:sro} shows sample configurations, computed specific heat capacity (SHC) and SRO parameters as a function of temperature for NiCoCr, NiCoFeCr, and NiCoFeMnCr. Configurations are visualised using OVITO~\cite{stukowski_visualization_2010}. We attribute the noise on the SHC curves to two causes. The first is that these quantities are obtained from Monte Carlo simulations and so a degree of noise is to be expected. The second is that for all of these systems there is no obvious single-phase ground state at the stoichiometries considered, and there will therefore be an element of frustration and disorder smearing in these curves. The peaks in the SHC curves, associated with a phase transition, are at lower values than predicted by the linear response analysis. This is to be expected however, as the linear response analysis represents a mean-field calculation and this is known to overestimate transition temperatures. We still find the trend that increasing configurational entropy lowers the ordering temperature, consistent with our earlier analysis.
\par
In all three systems, we see that the dominant correlations at high temperatures are a preponderance towards Co-Cr nearest neighbour pairs and away from both Ni-Co nearest neighbour pairs and Co-Cr second nearest neighbour pairs. This is in good agreement with both our linear response analysis and existing literature. For the NiCoCr system, previous studies have noted a tendency away from Cr-Cr nearest neighbour pairs and towards Co-Cr and Ni-Cr pairs, which is in good agreement with our findings~\cite{zhang_local_2017, ding_tunable_2018, zhang_short-range_2020}.
\par
Literature on the quarternary NiCoFeCr system is more limited, however one study by Tamm {\it et al} found that Cr-Cr pairs were disfavoured, while Ni-Cr and Co-Cr pairings were favoured, in agreement with our work~\cite{tamm_atomic-scale_2015}. The same work suggests that Cr atoms in this material begin to form a simple cubic lattice, which is consistent with our linear response analysis and existing literature~\cite{schonfeld_local_2019, niu_spin-driven_2015, fukushima_local_2017, he_understanding_2021}. However, Tamm {\it et al} also suggested that the number of Ni-Ni pairs decreases (we find an increase) and that there is significant SRO between Fe and the other three elements, while all our results show Fe interacting weakly and demonstrating little to no tendency to order. We believe that this disagreement arises because the work models the system in a ferromagnetic (FM) state, while we model the paramagnetic (PM) state. Fe in particular has a large local moment associated with it, and we expect the nature of predicted order in this material to be different between FM and PM states. Indeed, in an earlier work on Galfenol, a binary Fe-based alloy, we found that the nature of magnetic order significantly affected the nature of compositional order~\cite{marchant_ab_2021}. We only model the PM state in this work because the Curie temperature is around 120K~\cite{lucas_thermomagnetic_2013,billington_bulk_2020, schonfeld_local_2019}, well below the temperatures at which these materials are annealed~\cite{yin_yield_2020}. Another potential source of the discrepancy is that the Tamm study employed lattice MC simulations with energies evaluated {\it ab initio}, so was limited to less than 20 MC steps per atom, which they highlight as a potential limitation. By contrast, our extracted atomistic modelling parameters enable us to comfortably achieve $10^4$ MC steps per atom, which is the expected number needed to achieve equilibrium and enables us to interpret our results with confidence. We also note that another first principles study by Sch\"{o}nfeld {\it et al} found similar results to ours; Co-Cr and Ni-Cr correlations were dominant, with Fe exhibiting little to no correlation with other elements~\cite{schonfeld_local_2019}.
\par
Finally, for the quinary NiCoFeMnCr, there is little to no literature on SRO with which to compare. Our results suggest that any order in this material is likely to develop at such low temperatures as to be experimentally inaccessible; we find the pair distribution to be very close to uniform down to almost room temperature.
\par
For all three multicomponent systems at lower temperatures, we observe that the system forms a large number of Ni-Ni nearest neighbour pairs which, in combination with visual analysis of the generated configurations, we associate with the system phase segregating into an Ni-rich region and another region which is Ni-deficient. The temperature at which this occurs in our simulations is typically below room temperature so unlikely to be experimentally accessible, but nonetheless this weakens the case for any study searching for a single-phase, 0K ground state.

\subsection{Computational Resources}
At a time when the environmental cost of high-performance computing is under scrutiny~\cite{lannelongue_green_2021}, we consider it appropriate to summarise the computational resources used to obtain the data for this work and the trade off struck between computational cost and accurate information. A quick, back of the envelope calculation estimates that all data tabulated or visualised in this work can be reproduced in less than 300 CPU hours on the {\it Orac} cluster at the University of Warwick, which uses Intel E5-2680 v4 (Broadwell) processors. Self-consistent potentials and DoS calculations represent less than 5 CPU hours per material. The linear response code developed in-house takes a little longer, at around 40 CPU hours per material to generate the data necessary for eigenvalue plots,  transition temperatures/modes and atom-atom pair interactions but we emphasise here that main portion of the linear response calculations does not scale with the number of components considered. Finally, our atomistic model for MC simulations makes these calculations exceptionally cheap, and these each run on a single core for no more than a couple of hours.
\par
While we note that we have not included effects such as local lattice distortions, the low computational cost of our calculations combined with the successful testing of our results with existing literature suggests that our method is a highly efficient way of exploring the space of candidate multicomponent alloys, particularly when compared to high-throughput calculations on supercells using {\it ab initio} evaluations of energies, which use orders of magnitude more CPU hours.

\section{Summary and Conclusions}
\label{sec:conclusions}

We have presented results from of a first-principles theory~\cite{khan_statistical_2016} for the leading terms in an expansion of a Gibbs free energy of a multi-component alloy in terms of order parameters that characterize potential, compositional phases.  Using fairly modest computing resources, it describes atomic short-range order in homogeneously disordered phases and generates reliable pairwise interaction parameters suited for atomistic modelling in a multicomponent setting.  
In the case study of a subset of the Cantor-Wu alloys, NiCo, NiCoCr, NiCoFeCr and NiCoFeMnCr, our analysis demonstrates that, for these systems, it is clearly the configurational entropy that stabilises the solid solution and drives down any order/disorder transition temperature. We also find that it is consistently Co-Co, Co-Cr and Cr-Cr correlations which are strongest across all three multicomponent systems considered, and we have discussed the origin of this in terms of the electronic structure of the disordered alloy.

Atomistic modelling parameters for the systems have then been extracted and reveal the shortcomings of both the pseudobinary approximation and also  approximations that limit interactions to nearest-neighbour distance only. We have then used the pairwise parameters to perform lattice Monte Carlo simulations of NiCoCr, NiCoFeCr, and NiCoFeMnCr to elucidate the nature of compositional order in these materials. Consistent with most earlier works, we find that SRO emerges mainly in the form of Cr-Cr pairs being disfavored, while Co-Cr pairs are favored. For the five component system, we find little discernible SRO, which is consistent with the lack of published experimental measurements on this system.

Open questions remain about whether simple pairwise interactions such as those which we have used for our atomistic modelling are sufficient to describe systems which have developed significant partial order, \textit{e.g.} at low temperatures; here contributions from multi-site terms might be warranted. Comparison with supercell calculations of partially ordered alloys can be used to assess their validity and are in progress. 
We conclude by noting that the broad agreement with existing literature and experimental data for this group of alloys by our approach is very encouraging and supports its use as an effective, computationally fast way to study  multicomponent alloys. This is particularly so for those in which the nature of compositional order is disputed or not well-understood and also in the search for new alloy phases.

\begin{acknowledgments}
The present work formed part of the PRETAMAG project,
funded by the UK Engineering and Physical Sciences Research
Council, Grants No. EP/M028941/1, EP/W021331/1. C.D.W. is supported by a studentship within the UK Engineering and Physical Sciences Research
Council-supported Centre for Doctoral Training in Modelling of Heterogeneous Systems, Grant No. EP/S022848/1.
\end{acknowledgments}

\end{document}